\documentstyle[seceq,epsf]{ptptex}

\title{
A Parametric Representation of Critical Curves and Caustics 
for a Binary Gravitational Lens 
}

\author{
Hideki {\sc Asada}
\footnote{E-mail: asada@phys.hirosaki-u.ac.jp} 
}

\inst{
Institute for Astrophysics at Paris, 
98bis boulevard Arago, 75014 Paris, France\\
Faculty of Science and Technology, Hirosaki University, 
Hirosaki 036-8561, Japan
}

\recdate{
}

\abst{
We find a simple expression for critical curves of a binary
gravitational lens. On the basis of this, we present a parametric 
representation of such curves. The caustics can also be expressed 
with the same parameterization. 
The present result is helpful for efficiently constructing many 
templates of light curves due to binary systems, particularly 
extrasolar planets, which cause spikes in the light curves 
when a source crosses the caustics. 
}

\begin{document}

\maketitle

\section{Introduction}
Since the pioneering work of Schneider and Wei{\ss} (1986),\cite{SW} 
gravitational lensing by binary objects has attracted a great deal 
of interest. \cite{SEF}  It offers an important tool also in the
search for extrasolar planets,\cite{MP,GL,Bozza,Asada02a} 
since it is independent of the Doppler method, which cannot determine 
the planet mass because of the unknown inclination angle. \cite{MB}  
In general, critical curves, on which the lensing amplification 
diverges, play an important role in lensing studies: 
The number of images changes at these curves, and giant arcs are 
also observed (For a review, see Ref. 2)). 
To this time, the curves for binary gravitational lenses have been 
drawn by solving a lengthy equation. 
This is a fourth-order equation,\cite{ES,WP} which can be solved 
analytically only with a quite lengthy form of roots for instance 
by using Mathematica. \cite{Wolfram}
Thus it has been believed that no simple parametric representation 
can be obtained. 

It seems possible, however, to introduce a parameter using sine 
and cosine functions such that we can determine a simple 
representation of the curves, because neither the sine nor cosine 
function is algebraic. 
In the complex formalism, by contrast, a parameter is
introduced in the beginning, and the fourth-order equation is 
obtained at the end. \cite{WP} 
Here, we should recall that algebra allows only a finite number of 
operations of addition, subtraction, multiplication, division and 
root extractions. \cite{Waerden} 
However, introducing a parameter through sine and cosine 
functions changes the situation completely, because these 
functions themselves contain an infinite number of operations. 
Hence, our result is not algebraically equivalent to the previous one.  
The main purpose of this paper is to obtain a compact expression  
of the critical curves: We re-examine the equation for them 
and show that it can be rewritten in a tractable form. 
This is a key discovery, allowing us to parameterize the critical curves 
and caustics in a simple way, as shown in the following.

\section{Critical curves and caustics for a binary gravitational lens} 
\subsection{A lens equation}
We consider a binary system located a distance 
$D_{\mbox{L}}$ from the observer. 
It consists of two point masses, $M_1$ and $M_2$, 
whose positions are denoted by $\mbox{\boldmath $L_1$}$ and 
$\mbox{\boldmath $L_2$}$, respectively. 
The displacement vector from object 1 to 2 is written as 
$\mbox{\boldmath $L$}=\mbox{\boldmath $L_2$}-\mbox{\boldmath $L_1$}$. 
For later convenience, let us define the Einstein ring radius 
angle as 
\begin{equation}
\theta_{\mbox{E}}=
\sqrt{\frac{4GM D_{\mbox{LS}}}{c^2 D_{\mbox{L}} D_{\mbox{S}}}} ,  
\end{equation}
where $G$ is the gravitational constant, $M$ is the total mass, 
$M_1+M_2$, and $D_{\mbox{S}}$ and $D_{\mbox{LS}}$ denote the distances 
between the observer and the source and 
between the lens and the source, respectively. 
In units of the Einstein ring radius angle 
$\theta_{\mbox{E}}$, the lens equation reads 
\begin{equation}
\mbox{\boldmath $\beta$}=\mbox{\boldmath $\theta$}
-\Bigl( 
\nu_1 \frac{\mbox{\boldmath $\theta$}-\mbox{\boldmath $\ell$}_1}
{|\mbox{\boldmath $\theta$}-\mbox{\boldmath $\ell$}_1|^2} 
+\nu_2 \frac{\mbox{\boldmath $\theta$}-\mbox{\boldmath $\ell$}_2}
{|\mbox{\boldmath $\theta$}-\mbox{\boldmath $\ell$}_2|^2} 
\Bigr) , 
\label{lenseq}
\end{equation}
where $\mbox{\boldmath $\beta$}$ and $\mbox{\boldmath $\theta$}$ 
denote the vectors representing the positions of the source and image, 
respectively, and we defined the mass ratios $\nu_1$ and $\nu_2$ and 
the angular vectors for the positions and separation 
$\mbox{\boldmath $\ell$}_1$, $\mbox{\boldmath $\ell$}_2$ and 
$\mbox{\boldmath $\ell$}$ as 
\begin{eqnarray}
\nu_i&=&\frac{M_i}{M_1+M_2} , \\
\mbox{\boldmath $\ell$}_i&=&
\frac{\mbox{\boldmath $L$}_i}{D_{\mbox{L}}\theta_{\mbox{E}}} , \\
\mbox{\boldmath $\ell$}&=&
\frac{\mbox{\boldmath $L$}}{D_{\mbox{L}}\theta_{\mbox{E}}} , 
\end{eqnarray} 
for $i=1, 2$. 
Clearly, we have the identity $\nu_1+\nu_2=1$. 
Equation (\ref{lenseq}) is a set of two real coupled fifth-order 
equations for $(\theta_x, \theta_y)$, which is equivalent to 
a single complex quintic equation for 
$\theta_x+i \theta_y$. \cite{Witt90,Witt93} 
It has been shown recently that this equation can be reduced to 
a single real quintic equation. \cite{Asada02b} 

\subsection{Critical curves}
Critical curves are curves in the lens plane, where the Jacobian 
of the lens mapping 
$|\partial\mbox{\boldmath $\beta$}/\partial\mbox{\boldmath $\theta$}|$ 
vanishes. 
Now, we define the angular separation between the image and 
the lens $i$ ($i=1,2$) as 
\begin{equation}
\rho_i=|\mbox{\boldmath $\theta$}-\mbox{\boldmath $\ell$}_i| ,  
\end{equation}
where $\rho_i$ does not vanish, because the location of the image is 
different from those of the lens objects. 

Let us denote the height of the image measured from 
the separation axis by $h$. 
Then, straightforward computations of a vanishing Jacobian give us 
\begin{equation}
\rho_1^4\rho_2^4=(\nu_2\rho_1^2+\nu_1\rho_2^2)^2
-4\nu_1\nu_2\ell^2h^2 . 
\label{critical0}
\end{equation}
By rearranging the expression for the separation as 
$\ell=\sqrt{\rho_1^2-h^2}+\sqrt{\rho_2^2-h^2}$, 
we find 
\begin{equation}
h^2=\frac{4\rho_1^2\rho_2^2-(\ell^2-\rho_1^2-\rho_2^2)^2}{4\ell^2} . 
\end{equation}
Substituting this into Eq. ($\ref{critical0}$), we obtain 
\begin{equation}
\rho_1^4\rho_2^4=\nu_2(\rho_1^2-\nu_1 \ell^2)^2
+\nu_1(\rho_2^2-\nu_2 \ell^2)^2 , 
\label{critical1}
\end{equation}
where we defined $\ell=|\mbox{\boldmath $\ell$}|$. 
Here, it should be noted that this expression is independent of 
the choice of the origin of our coordinates. 
In the complex formalism, for instance, the equation for 
the critical curves is inevitably a fourth-order polynomial equation, 
whose coefficients depend on the choice of the origin of the
coordinates, \cite{WP} so that we need a numerical method to solve 
the quartic, and the resultant expressions are quite lengthy. 

Our procedure for obtaining a parametric expression of critical 
curves is divided into two steps. 
First, we find a simple parametric expression of 
$\rho_1$ and $\rho_2$. Next, using this expression, 
we obtain those of $\theta_x$ and $\theta_y$. 
For convenience, we define 
\begin{eqnarray}
u&=&\frac{\rho_1^2}{\sqrt{\nu_1}} , 
\label{u} \\
v&=&\frac{\rho_2^2}{\sqrt{\nu_2}} , 
\label{v}
\end{eqnarray}
so that Eq. ($\ref{critical1}$) is rewritten simply as
\begin{equation}
1=\frac{\left(1-\frac{\sqrt{\nu_1}\ell^2}{u}\right)^2}{v^2}
+\frac{\left(1-\frac{\sqrt{\nu_2}\ell^2}{v}\right)^2}{u^2} , 
\end{equation}
which allows us to introduce a parameter $s \in [0,2\pi)$ 
according to the relations  
\begin{eqnarray}
\cos s&=&\frac{1-\frac{\sqrt{\nu_1}\ell^2}{u}}{v} , 
\label{cos}\\
\sin s&=&\frac{1-\frac{\sqrt{\nu_2}\ell^2}{v}}{u} . 
\label{sin}
\end{eqnarray}
Equation ($\ref{cos}$) can be rewritten as 
\begin{equation}
\frac1v=\frac{\cos s}{1-\frac{\sqrt{\nu_1}l^2}{u}} . 
\end{equation}
Substitution of this into Eq. ($\ref{sin}$) gives us 
a quadratic equation for $u$:  
\begin{equation}
(\sin s)u^2-(\sqrt{\nu_1}\ell^2\sin s-\sqrt{\nu_2}\ell^2\cos s+1)u
+\sqrt{\nu_1}\ell^2=0 . 
\label{eq-u}
\end{equation}
Equation ($\ref{eq-u}$) is solved as 
\begin{equation}
u=\frac{A\pm\sqrt{B}}{2\sin s} , 
\label{u2}
\end{equation}
where we defined 
\begin{eqnarray}
\cos\alpha&=&\sqrt{\nu_1} , \\
\sin\alpha&=&\sqrt{\nu_2} , \\
A&=&1+\ell^2\sin(s-\alpha) , \\
B&=&A^2-4\ell^2\cos\alpha\sin s . 
\end{eqnarray}
Substituting Eq. ($\ref{u2}$) into $u$ of Eq. ($\ref{cos}$) leads to 
\begin{equation}
v=\frac{C\pm\sqrt{B}}{2 \cos s} ,
\label{v2}
\end{equation}
where we defined 
\begin{equation}
C=1-\ell^2\sin(s-\alpha) . 
\end{equation}

Up to this point, we have not specified the coordinate system. 
Now, let us introduce polar coordinates $(r, \phi)$ whose origin 
is located at the mass $M_1$ and whose angle is measured from 
the separation axis of the binary. 
This implies that $\mbox{\boldmath $\ell$}_1=0$ and 
$\mbox{\boldmath $\ell$}_2=\mbox{\boldmath $\ell$}$. 
In terms of the polar coordinates, $\rho_1$ and $\rho_2$ 
are expressed as 
\begin{eqnarray}
\rho_1^2&=&r^2 , 
\label{rho1} \\
\rho_2^2&=&r^2-2\ell r\cos\phi+l^2 . 
\label{rho2} 
\end{eqnarray}
Then, Eqs. ($\ref{rho1}$) and ($\ref{rho2}$) are solved for $r$ 
and $\cos\phi$ as 
\begin{eqnarray}
r&=&\nu_1^{1/4}\sqrt{u} , \\
\cos\phi&=&
\frac{\sqrt{\nu_1}u-\sqrt{\nu_2}v+\ell^2}{2\ell\nu_1^{1/4}\sqrt{u}} , 
\end{eqnarray}
where we used Eqs. ($\ref{u}$), ($\ref{v}$), ($\ref{u2}$) 
and ($\ref{v2}$). 
Hence, in terms of $u$ and $v$, the critical curves are expressed as 
\begin{eqnarray}
\theta_x&=&r\cos\phi 
\nonumber\\
&=&\frac{\sqrt{\nu_1}u-\sqrt{\nu_2}v+\ell^2}{2\ell} , 
\label{thetax} \\
\theta_y&=&\pm r\sqrt{1-\cos^2\phi} 
\nonumber\\
&=&\pm\frac{\sqrt{D}}{2\ell} , 
\label{thetay} 
\end{eqnarray}
where we defined 
\begin{equation}
D=-\nu_1u^2-\nu_2v^2-\ell^4+2\sqrt{\nu_1}\ell^2u+2\sqrt{\nu_2}\ell^2v
+2\sqrt{\nu_1\nu_2}uv . 
\end{equation}

We are now in a position to consider the constraint on $\rho_1$ 
and $\rho_2$, that the two lens objects and the image form a triangle. 
This condition is expressed as 
\begin{equation}
|\rho_1-\rho_2|<\ell , 
\label{triangle1}
\end{equation}
which can be rewritten as 
\begin{equation}
|\nu_1^{1/4}\sqrt{u}-\nu_2^{1/4}\sqrt{v}|<\ell . 
\label{triangle2}
\end{equation}
Since $u$ and $v$ are positive, we can show that this inequality is 
equivalent to the relation $D>0$. 

It is possible in principle to determine allowed regions for $s$ 
by analytically treating Eq. ($\ref{triangle2}$), 
which is the condition for a triangle. 
In practice, however, it is much easier to draw the critical curves 
for $s \in [0, 2\pi)$ only if we ignore $s$ for $D<0$. 

In the case of a symmetric binary, i.e. $\nu_1=\nu_2=1/2$, 
the condition $D>0$ becomes 
\begin{equation}
(\sin s+\cos s)^2+2\sqrt{2}(\sin s+\cos s)-4<0 . 
\end{equation}
This is solved as 
\begin{equation}
\sin s+\cos s<\sqrt{6}-\sqrt{2} , 
\end{equation}
where we used the fact that both $|\sin s|$ and $|\cos s|$ are 
less than unity.
Hence, we find 
\begin{equation}
\frac{3\pi}{4}-\sin^{-1}(\sqrt3-1)<u<\frac{7\pi}{4}
+\sin^{-1}(\sqrt3-1) , 
\end{equation}
where we assumed $s \in [\pi/2, 5\pi/2)$ for convenience. 
Otherwise, for $[0, 2\pi)$, the allowed regions are apparently 
divided into two, $s \in [0, \sin^{-1}(\sqrt3-1)-\pi/4)$ 
and $[3\pi/4-\sin^{-1}(\sqrt3-1), 2\pi)$. 

\begin{figure}
\epsfxsize 9cm
\begin{center}
\epsfbox{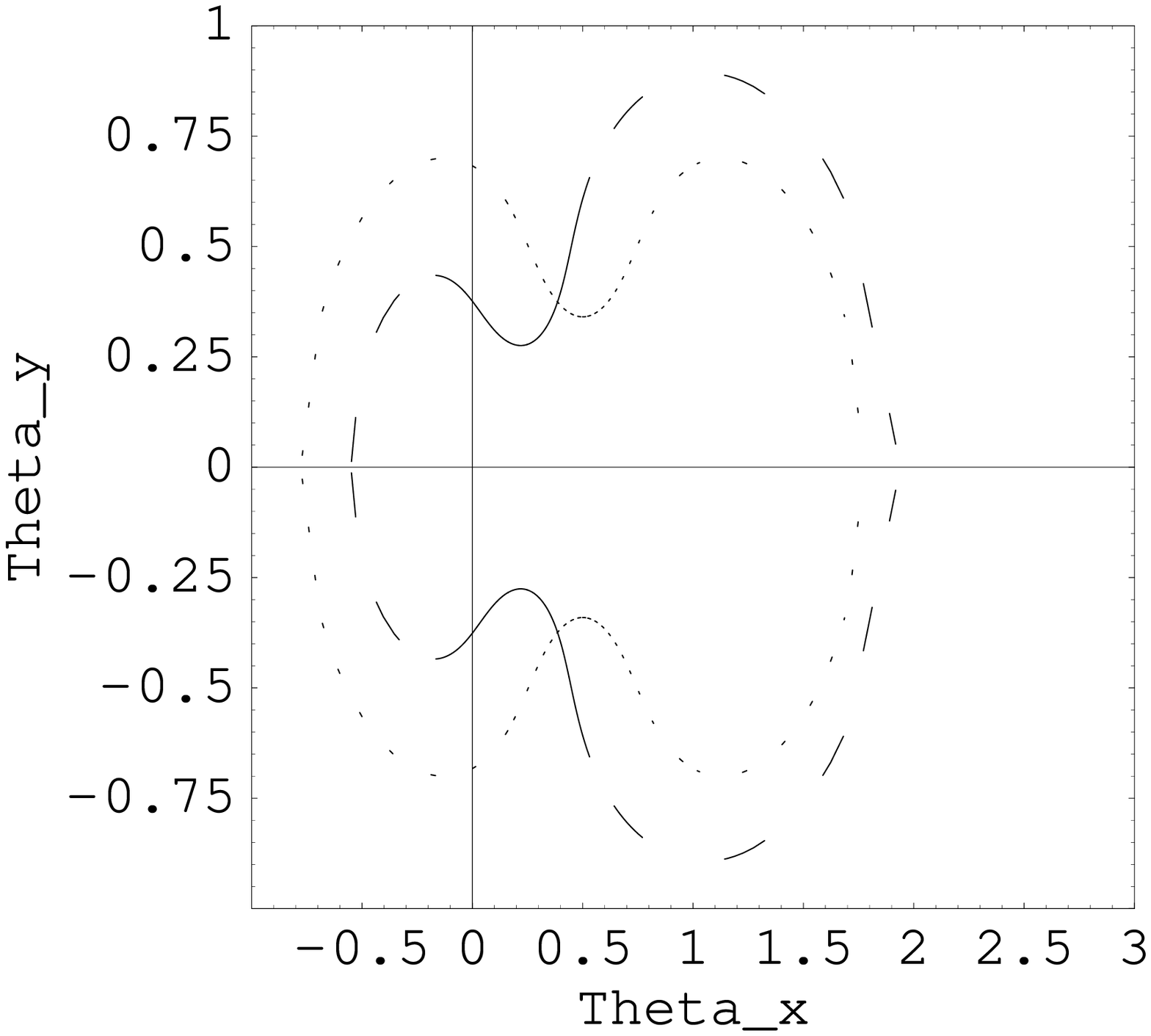}
\end{center}
\caption{Critical curves for a binary gravitational lens 
with the separation of the Einstein ring radius $\ell=1$. 
The dotted and dashed curves correspond to the mass ratios 
$\nu_1=1/2$ and $1/5$, respectively. 
}
\label{Fig1}
\end{figure}

\begin{figure}
\epsfxsize 9cm
\begin{center}
\epsfbox{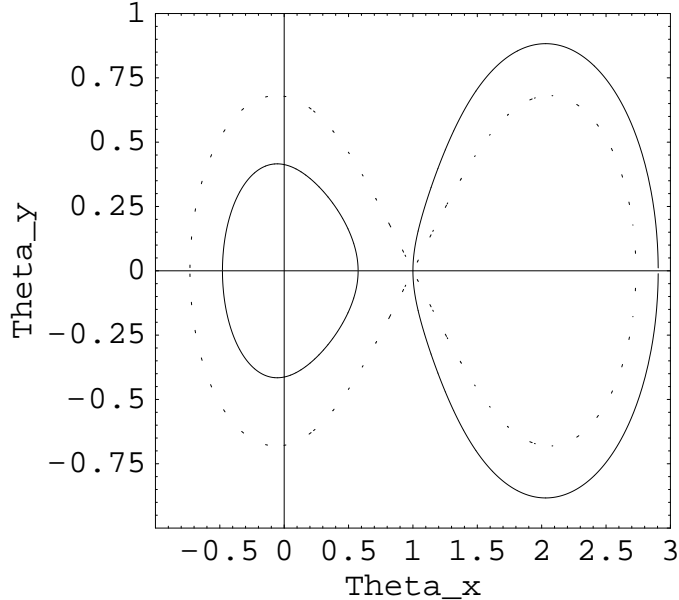}
\end{center}
      \caption{Critical curves for a binary gravitational lens 
with the separation of the Einstein ring diameter $\ell=2$. 
The dotted and solid curves correspond to the mass ratios 
$\nu_1=1/2$ and $1/5$, respectively. 
}
\label{Fig2}
\end{figure}

As an illustration, critical curves are given for two cases, 
that in which the binary separation equals the Einstein ring radius 
(Fig.1), and that in which the separation equals the Einstein 
ring diameter (Fig.2). 
These figures demonstrate that the value of the binary separation is 
of crucial importance in determining topological properties of the 
critical curves, in particular, whether there is one loop or two, 
and that a non-half mass ratio causes the curves to be asymmetric. 
\cite{SW,WP}

\subsection{Caustics}
The parametric representation of critical curves is 
given by Eqs. ($\ref{thetax}$) and ($\ref{thetay}$). 
Substitution of these into the lens equation ($\ref{lenseq}$) 
with $\mbox{\boldmath $\ell$}_1=0$ and 
$\mbox{\boldmath $\ell$}_2=\mbox{\boldmath $\ell$}$ gives immediately 
a representation of caustics in terms of the parameter $s$, 
\begin{equation}
\mbox{\boldmath $\beta$}(s)=\mbox{\boldmath $\theta$}(s)
-\Bigl( 
\nu_1 \frac{\mbox{\boldmath $\theta$}(s)}{|\mbox{\boldmath $\theta$}(s)|^2} 
+\nu_2 \frac{\mbox{\boldmath $\theta$}(s)
-\mbox{\boldmath $\ell$}}{|\mbox{\boldmath $\theta$}(s)
-\mbox{\boldmath $\ell$}|^2} 
\Bigr) , 
\label{caustics}
\end{equation}
where $\mbox{\boldmath $\theta$}(s)$ is given by 
Eqs. ($\ref{thetax}$) and ($\ref{thetay}$). 

\begin{figure}
\epsfxsize 9cm
\begin{center}
\epsfbox{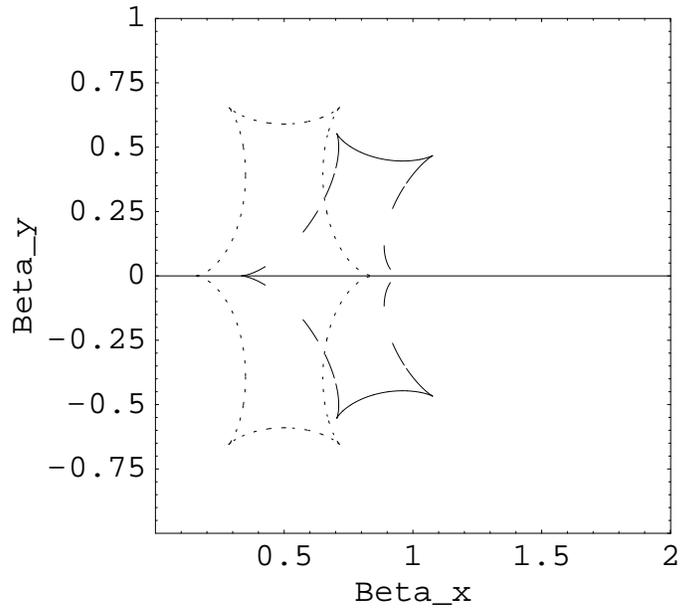}
\end{center}
      \caption{Caustics for a binary gravitational lens 
with the separation of the Einstein ring radius $\ell=1$. 
Each curve corresponds to that in Fig. 1. 
}
\label{Fig3}
\end{figure}
\begin{figure}
\epsfxsize 9cm
\begin{center}
\epsfbox{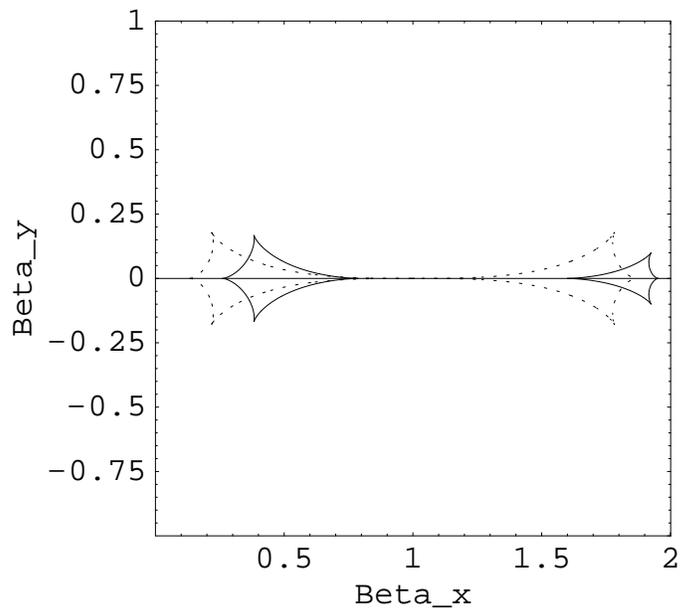}
\end{center}
      \caption{Caustics for a binary gravitational lens 
with the separation of the Einstein ring diameter $\ell=2$. 
Each curve corresponds to that in Fig. 2. 
}
\label{Fig4}
\end{figure}

The caustics corresponding to Figs. 1 and 2 are given in 
Figs. 3 and 4, which show that the binary separation is important 
in determining the topological features of caustics, while caustics 
are distorted due to the mass ratio.

\section{Conclusion}
First, an equation for the critical curves of a binary gravitational 
lens was recast into the compact form ($\ref{critical1}$).  
On the basis of this, we obtained a parametric representation 
of these curves as Eqs.($\ref{thetax}$) and ($\ref{thetay}$). 
A parametric representation of the caustics was also obtained as 
Eq. ($\ref{caustics}$). These representations are helpful 
to study binary lensing, particularly for a number of sets of 
the mass ratio and separation. 

For instance, when we discover extrasolar planets by microlensing, 
a standard strategy is to detect a slight change in light curves 
using a small telescope, and then to follow this up using 
a large telescope on the ground or in space as soon as possible. 
In order to be able to respond as quickly as possible, it is necessary 
to prepare many light curve templates, in which spikes appear 
owing to planetary microlensing. The positions of such spikes, 
which correspond to crossing caustics, are essential for templates 
used in initial observations. The detailed shapes of the spikes are
important to allow for analysis of light curves with follow-up 
observations. The expression obtained in this paper may be helpful 
for constructing these templates with good accuracy and efficiency. 
This is a future subject.

\section*{Acknowledgements}
The author would like to thank M. Kasai and T. Kasai 
for useful conversations, and N. Seymour for carefully reading 
the revised manuscript. 
He would like to thank L. Blanchet for his hospitality at the 
institute for Astrophysics at Paris, where this work was done. 
This work was supported by a fellowship for visiting scholars from 
the Ministry of Education of Japan.


\end{document}